\documentclass[
superscriptaddress,
 amsmath,amssymb,
 aps,
 pra,
 twocolumn,
 10pt,
 longbibliography
]{revtex4-2}

\usepackage{algorithm}
\usepackage[noend]{algpseudocode}

\usepackage{orcidlink}
\usepackage[utf8]{inputenc}
\usepackage[T1]{fontenc}
\usepackage{graphicx}
\usepackage{array}
\usepackage{braket}

\usepackage{booktabs}
\usepackage{multirow}
\usepackage{color,soul}
\usepackage{mathtools}
\usepackage{hyperref}
\hypersetup{
	colorlinks,
	linkcolor={blue},
	citecolor={blue},
	urlcolor={blue}
}

\usepackage{mdframed}

\newcounter{algobox}
\renewcommand{\thealgobox}{\arabic{algobox}}

\newenvironment{algobox}[1][]{
  \refstepcounter{algobox}
  \begin{center}
    Box \thealgobox. \textbf{#1}.
  \end{center}
  \begin{mdframed}[linewidth=0.8pt,roundcorner=5pt,
                   innerleftmargin=1pt,innerrightmargin=1pt,
                   innertopmargin=1pt,innerbottommargin=1pt]
}{
  \end{mdframed}
}

\begin{document}

\title{Local Clustering Decoder as a fast and adaptive hardware decoder for the surface code}

\author{Abbas B. Ziad\,\orcidlink{0009-0008-2517-0619}}
\email{abbasbrackenziad@gmail.com}
\affiliation{Riverlane, Cambridge, CB2 3BZ, UK}
\author{Ankit Zalawadiya\,\orcidlink{0009-0008-4399-5892}}
\affiliation{Riverlane, Cambridge, CB2 3BZ, UK}
\author{Canberk Topal\,\orcidlink{0009-0005-8933-8896}}
\affiliation{Riverlane, Cambridge, CB2 3BZ, UK}
\author{Joan Camps\,\orcidlink{0000-0003-3896-7922}}
\email{joan.camps@riverlane.com}
\affiliation{Riverlane, Cambridge, CB2 3BZ, UK}
\author{Gy\"{o}rgy P. Geh\'er\,\orcidlink{0000-0003-1499-3229}}
\affiliation{Riverlane, Cambridge, CB2 3BZ, UK}
\author{Matthew P. Stafford\,\orcidlink{0000-0001-9938-3854}} 
\affiliation{Riverlane, Cambridge, CB2 3BZ, UK}
\affiliation{Quantum Engineering Technology Labs, H. H. Wills Physics Laboratory and Department of Electrical \& Electronic Engineering, University of Bristol, BS8 1FD, UK}
\author{Mark L. Turner\,\orcidlink{0000-0002-5274-957X}}
\affiliation{Riverlane, Cambridge, CB2 3BZ, UK}
\date{August 2025}

\begin{abstract}
To avoid prohibitive overheads in performing fault-tolerant quantum computation, the decoding problem needs to be solved accurately and at speeds sufficient for fast feedback. Existing decoding systems fail to satisfy both of these requirements, meaning they either slow down the quantum computer or reduce the number of operations that can be performed before the quantum information is corrupted. We introduce the Local Clustering Decoder as a solution that simultaneously achieves the accuracy and speed requirements of a real-time decoding system. Our decoder is implemented on FPGAs and exploits hardware parallelism to keep pace with the fastest qubit types. Further, it comprises an adaptivity engine that allows the decoder to update itself in real-time in response to control signals, such as heralded leakage events. Under a realistic circuit-level noise model where leakage is a dominant error source, our decoder enables one million error-free quantum operations with 4x fewer physical qubits when compared to standard non-adaptive decoding. This is achieved whilst decoding in under 1~$\mu$s per round with modest FPGA resources, demonstrating that high-accuracy real-time decoding is possible, and reducing the qubit counts required for large-scale fault-tolerant quantum computation.
\end{abstract}

\maketitle

\section{Introduction}
To realise the potential of quantum computation, where useful quantum algorithms outperform the best classical approaches \cite{gidney_how_2021, lee_even_2021}, devices must be able to operate in a large-scale, fault-tolerant regime. In response to qubit fragility, Quantum Error Correction (QEC) has emerged as a fundamental enabling technology. Provided the physical error rate of the system is below a threshold, QEC techniques can leverage a polynomial overhead in resources to redundantly encode quantum information, exponentially suppressing logical errors \cite{aharonov_fault-tolerant_1999}.

A key element of a fault-tolerant quantum computer is the decoder. Decoders process a stream of data (rounds of error ``syndromes'') to infer the logical consequences of the physical noise affecting the system. Good decoders must be accurate and have high throughput and low latency. Accuracy is needed to avoid excessive QEC resource overheads. High throughput is essential to avoid the backlog problem, where syndromes accumulate faster than the decoder can process \cite{terhal_quantum_2015}. Low latency enables quick response times, which are needed for fast logical clock rates. On some qubit platforms, a single round of syndrome extraction takes around 1~$\mu$s, requiring MHz throughput and latencies in the few tens of $\mu$s. These demanding latency and throughput requirements have been met in implementations on dedicated classical hardware, such as Field Programmable Gate Arrays (FPGAs) and Application Specific Integrated Circuits (ASICs)---but, so far, at a cost in accuracy \cite{overwater_neural-network_2022, liao_wit-greedy_2023, vittal_astrea_2023, liyanage_scalable_2023, barber_real-time_2023, liyanage_fpga-based_2024} or scalability \cite{das_lilliput_2022}. Real-time software decoders have been shown to deliver high throughput and high accuracy---but at a cost in latency~\cite{acharya_quantum_2024}.

Here, we introduce the Local Clustering Decoder (LCD): a surface code decoder that retains the performance advantage offered by hardware decoders, while obtaining levels of accuracy and flexibility that are competitive with their software counterparts.
LCD is an adaptive and distributed version of an error-clustering algorithm based on Union-Find \cite{delfosse_almost-linear_2021}, that we implement on FPGAs. It consists of two key innovations over previous hardware decoders. Firstly, we use a coarse-grained parallel architecture that results in lower FPGA resource utilisation when compared to fine-grained parallel architectures.
Secondly, LCD can update its error model on the fly using a set of pre-learned adaptations computed by an adaptivity engine. Together, these innovations allow LCD to deliver high accuracy without compromising on the hardware decoder performance. 

We demonstrate the benefits of LCD by decoding a surface code patch under a circuit-level noise model with leakage---a damaging correlated noise channel affecting most qubit types \cite{suchara_leakage_2014, brown_handling_2019}. We observe a significant improvement in the error-correction performance when leakage adaptivity is included, effectively halving the code distance $d$ required for computation: in our noise model, from $d=33$ to $d=17$ for one million error-free logical operations. Note that this leads to a $75\%$ qubit saving compared to the requirements of a non-adaptive decoder.  At $d=17$ our decoder uses less than 10\% of the available resources of a high-end Xilinx FPGA, whilst being comfortably within the time budget of 1 $\mu\textrm{s}$ per round of syndrome extraction. We foresee the adaptivity engine facilitating further improvements in accuracy via, e.g., two-pass correlated decoding \cite{fowler_optimal_2013}. Higher accuracy will lead to further qubit savings.

\section{Results}
\subsection{Decoding the Surface Code}
\begin{figure}[t!]
\centering
\includegraphics[clip, trim=1cm 8.7cm 0cm 5cm, width=0.556\textwidth]{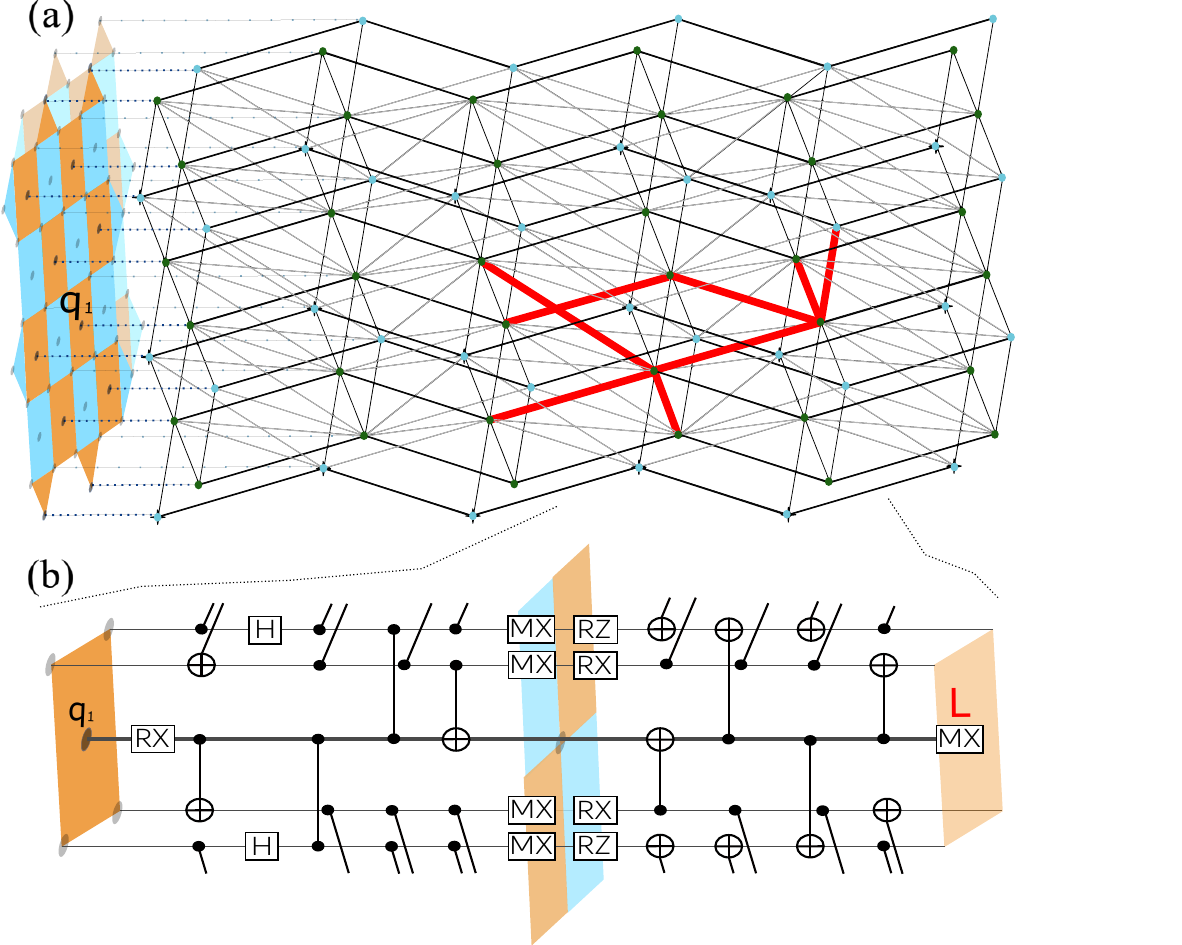}
\caption{\textbf{Circuitry and decoding graphs.} 
{\bf (a)} A distance $5$ rotated planar surface code and its associated $Z$-type decoding graph. $X$ ($Z$) plaquettes are orange (blue). Light-blue vertices share an edge with a virtual boundary vertex (not shown). This edge flips a logical observable if the vertex is marked with a cross. {\bf (b)} Circuitry implementing patch wiggling over two rounds of syndrome extraction. Notice that hook edges in the decoding graph reverse every round. This is due to reversing the scheduling of the stabilisers, which is inherently needed to achieve wiggling \cite{mcewen_relaxing_2023, geher_error-corrected_2023}. In red, we show the set of errors made more likely when the qubit labelled $q_1$ is measured as leaked at the end of the fourth round of syndrome extraction. These errors follow the 2-qubit gates involving $q_1$ depicted in the circuit.}
\label{fig:wiggling}
\end{figure}

The surface code is a leading candidate for achieving fault tolerance at scale \cite{kitaev_fault-tolerant_2003} since it can be implemented on hardware with fixed two-dimensional qubit connectivity, such as superconducting platforms---where QEC with the surface code has been demonstrated experimentally \cite{marques_logical-qubit_2022, krinner_realizing_2022, acharya_suppressing_2023, acharya_quantum_2024}. A (rotated) surface code patch of size $d\times d$ requires $2d^2-1$ physical qubits and encodes a single logical qubit. Fig.~\ref{fig:wiggling}~(a) shows a distance 5 surface code.   Physical qubits are divided into two roles: data qubits, carrying logical information, are located on plaquette vertices; auxiliary qubits, detecting errors, are located at the centre (on the apex) of bulk (boundary) plaquettes.

The simplest approach to decoding the surface code uses a graph approximation to the errors in the system that treats $X$ and $Z$ errors separately, and decomposes $Y$ errors into the product of independent $X$ and $Z$ errors \cite{dennis_topological_2002, fowler_minimum_2014, higgott_pymatching_2022, delfosse_almost-linear_2021}. In this decoding graph, error mechanisms are represented by edges triggering ``detectors'' at the vertices. Errors triggering only one detector are represented by edges connecting this detector to an auxiliary ``boundary vertex''. This graph approximation enables two important decoding strategies: matching pairs of triggered detectors by short edge chains of errors; and finding small clusters with even numbers of triggered detectors, for example using Union-Find \cite{delfosse_almost-linear_2021}. By adding weights to the edges of the graph, we can represent varying prior probabilities of distinct errors.

Below threshold, the logical error rate behaves as 
\begin{equation}
\label{eq:lambda}
P_L \propto \Lambda^{-d/2}\,
\end{equation}
where $\Lambda>1$ and $d$ is the distance of the code---logical errors are exponentially suppressed. On a fixed device, less accurate decoders offer smaller $\Lambda$, requiring a larger $d$, and so more qubits, to achieve a desired target logical error rate $P_L$. By updating a graph error model on the fly, graph-based decoders can deliver an enhanced $\Lambda$. For example, $Y$ errors can be decoded more accurately when the weights of the $X$ ($Z$) graph are updated conditioned on the outcome of decoding in $Z$ ($X$) \cite{acharya_quantum_2024}, substantially increasing $\Lambda$ \cite{fowler_optimal_2013}. We call this capability of real-time modification of a decoding graph ``adaptivity''. 

Another type of noise which greatly benefits from adaptive decoding is leakage \cite{fowler_coping_2013, suchara_leakage_2014, brown_critical_2020}. Leakage is a particularly damaging, long-lived error mechanism where a qubit exits the $\{\ket{0}, \ket{1}\}$ computational subspace in an uncontrolled manner. Though leakage events may happen rarely, when a qubit has leaked it will damage the qubits it interacts with until it returns to the computational subspace \cite{aliferis_fault-tolerant_2007, fowler_coping_2013, suchara_leakage_2014, brown_leakage_2019, brown_handling_2019, brown_critical_2020}. Therefore a single leakage event can introduce large amounts of subsequent noise into the system. With no mechanism to restore qubits to their computational subspace, a single leakage event can introduce a logical failure, effectively destroying the distance property of the code, and negating the exponential suppression of Eq.~\eqref{eq:lambda} \cite{aliferis_fault-tolerant_2007, fowler_coping_2013}.

Leakage Reduction Units (LRUs) are gadgets designed to return leaked qubits to the computational subspace, restoring Eq.~\eqref{eq:lambda}. LRUs must be used for implementing QEC in the presence of leakage \cite{aliferis_fault-tolerant_2007}. A simple way to return a leaked qubit to the computational subspace is to perform a multi-level reset \cite{mcewen_removing_2021}. Conventionally, auxiliary qubits are measured and reset every round. Data qubits, on the other hand, are conventionally not frequently reset. Here we employ an LRU called patch-wiggling \cite{mcewen_relaxing_2023}, which alternates the role of data/auxiliary qubits each round, so that every qubit is reset every two rounds. This limits the lifetime of leakage to two rounds of syndrome extraction. As shown in Fig.~\ref{fig:wiggling}~(a), wiggling changes the structure of the decoding graph compared with the standard QEC circuit.

If leakage can be resolved at measurement (i.e., heralded), we know that the qubit must have leaked at some point over its two-round lifetime \cite{suchara_leakage_2014}. Through LCD's adaptivity engine, we can use this heralding information to update our prior by modifying the decoding graph (see Methods). In this work, we do so by initiating the decoder with preclusters of likely errors around the leaked qubit measurements. Fig.~\ref{fig:wiggling}(a) shows an example modification where qubit $q_1$ has leaked in the fourth round of syndrome extraction. The edges of the graph to be modified upon heralding the leakage event are highlighted in red.  As we will see, this greatly improves the accuracy of our decoder to $\Lambda^2$---halving the distance required to attain a target performance when leakage is a dominant noise source. A full description of our leakage model is provided in Methods. 

\begin{figure*}[ht]
\centering
\includegraphics[scale=0.41]{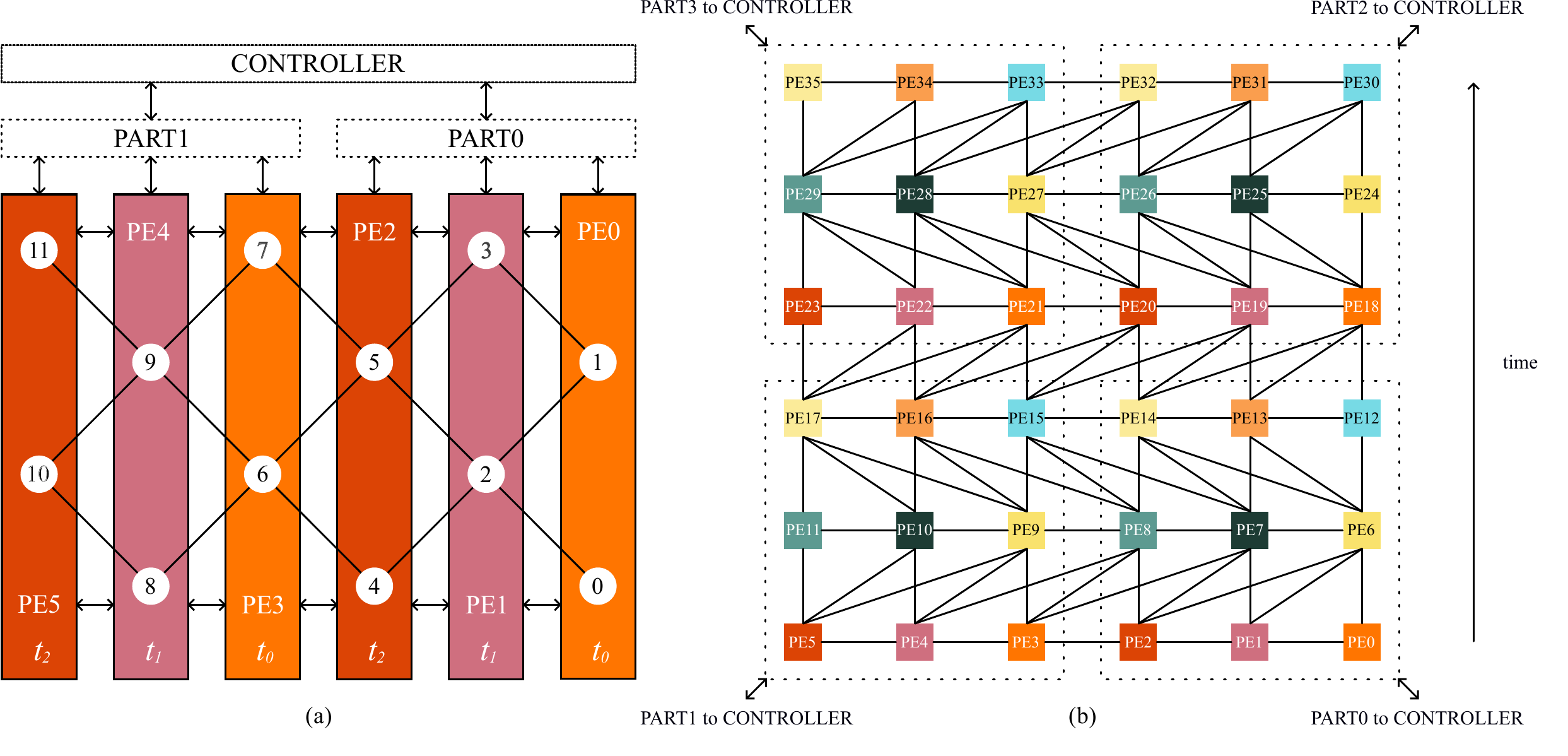}
\caption{\textbf{Example decoding engine. (a)} A single round of the $d = 5$ rotated planar surface code mapped onto a PE array. Each PE is assigned two vertices and linked to the PEs that contain their neighbours. Furthermore, each PE is assigned to a time slot in either the first or second part of a pair of conflict-free parts. The PEs in the same colour class are in the same time slot. The central controller coordinates the parts, which in turn coordinate the PEs. \textbf{(b)} The decoding engine extended in time to include $d + 1$ contiguously stacked layers, capturing graphlike error mechanisms across $d$ rounds of syndrome extraction. There are now four parts, each containing nine PEs. Note that the assignment of PEs to time slots is equivalent to a colouring of the PEs in the square of the array---the square of the array is constructed by adding links between any two PEs whose distance in the array is 2. The straight links connecting PEs in the same row/column support the spatial/timelike edges in the decoding graph and the diagonal links connecting PEs one/two columns apart support the short/long hook edges. Note that the diagonal links reverse direction every round. This is done to support patch wiggling, cf.~Fig.~\ref{fig:wiggling}. In general, the PE array is compiled from the decoding graph to support arbitrary shapes and sizes.}
\label{fig:decoding_engine}
\end{figure*}

\subsection{Local clustering decoder (LCD)}

Our decoder is composed of two main components: (1) a decoding engine that projects the decoding graph onto a coarse-grained parallel architecture and implements a distributed clustering algorithm based on Union-Find to achieve an average per round decoding time that scales sublinearly with the surface code distance; and (2) an adaptivity engine that uses a pre-learned set of adaptions to update the decoding graph at runtime in response to control signals.

\subsubsection{Decoding engine}

Parallel algorithms for decoding surface codes have been extensively studied in the literature, starting with Fowler's minimum-weight perfect matching (MWPM) proposal \cite{fowler_minimum_2014} and culminating in a series of realisations based on Union-Find \cite{liyanage_scalable_2023, chan_actis_2023, heer_achieving_2023, liyanage_fpga-based_2024}. In these schemes, the decoding graph is mapped onto an array of processing elements (PEs) and the decoding problem is solved in a distributed fashion to maximise throughput.

In a fine-grained mapping, each PE is assigned a single vertex in the decoding graph \cite{liyanage_scalable_2023, chan_actis_2023}; in a coarse-grained mapping, each PE is assigned multiple vertices \cite{heer_achieving_2023, liyanage_fpga-based_2024}. The former offer the highest throughput at the cost of high resource utilisation. The latter assign more work to each PE in an attempt to reduce resource utilisation. Since each PE takes longer to complete a task, this results in a lower throughput, but this is typically tolerable within the bounds of the superconducting QEC cycle time (1~$\mu$s).  

Our decoding engine aims to balance the space-time tradeoff by selecting a level of granularity that results in high throughput and modest resource utilisation. For example, Fig.~\ref{fig:decoding_engine}~(a) shows how we map a single layer of the decoding graph for the distance 5 rotated planar surface code onto a PE array. Each PE is assigned a batch of vertices, which it stores in a local memory. If a vertex in one PE is adjacent to a vertex in another, the PEs are connected via bidirectional links.

To avoid contention, we require that each PE be in at most one transaction on either the sending/receiving side. This can be ensured by partitioning the PE array into a set of pairwise conflict-free parts. A part is a collection of PEs each assigned to a unique time slot, and a pair of parts is conflict-free if the PE assigned to slot $t_i$ of the first is at least three links from the PE assigned to slot $t_i$ of the second. For example, the pair of parts in Fig.~\ref{fig:decoding_engine} is conflict-free, since the PEs assigned to time slots $t_0$, $t_1$ and $t_2$ are all exactly three links from each other. If a stage requires inter-PE communication, each part must execute its PEs serially, one slot at a time; however, the stage logic can be shared amongst the PEs to save resources. Otherwise, each part may execute its PEs in parallel, provided the stage logic is replicated in each of them. A central controller coordinates the parts, which in turn coordinate the PEs.

Fig.~\ref{fig:decoding_engine}~(b) shows how the decoding engine in Fig.~\ref{fig:decoding_engine}~(a) is extended to support $d = 5$ measurement rounds. In contrast to the three-dimensional structures inherent in fine-grain architectures, our decoding engine can be laid out in two-dimensions. This makes it easier to place and route on silicon platforms, such as FPGAs and ASICs. Here, we also demonstrate how our decoding engine is compiled for intricate decoding graphs by appropriately placing the inter-PE links. Specifically, patch wiggling is enabled simply by reversing the direction of the hook links every round.

Typically, experimentalists will choose software over hardware decoders due to their flexibility \cite{acharya_quantum_2024}. This is natural since QEC is a new field and the most optimal circuits are still being discovered. For example, syndrome extraction circuits can be structured to mitigate fabrication defects \cite{debroy_luci_2024}, implemented with iSWAP instead of CNOT or CZ gates \cite{mcewen_relaxing_2023}, with or without resets \cite{geher_reset_2024}, and various other constructions to improve logical error rates. However, each of these choices have implications on the decoding graph and therefore the problem being solved by the decoder. 

Our decoding engine can adjust to this evolving landscape since it is compiled from the decoding graph. Through an in-house framework, we compile the circuit defined via Stim's domain-specific language (DSL) \cite{gidney_stim_2021} down to an adjacency graph representation of the decoding problem. The adjacency graph is then used to auto-generate the SystemVerilog Network-on-Chip (NoC) module that defines the inter-PE links. This infrastructure enables us to bring-up and validate our decoder against new experiments in timelines that are competitive with software decoders.

\subsubsection{Adaptivity engine}

We define adaptivity as the ability of a decoder to change properties associated with the edges in its decoding graph at runtime. This facilitates tackling correlated noise channels such as leakage \cite{suchara_leakage_2014}.

To enable real-time adaptivity with minimal latencies, we require adaptations to be precomputed. This means that updates can be applied efficiently on the fly as small, targeted modifications to a minimal subset of the edges in the decoding graph. Accordingly, LCD's adaptivity engine is responsible for storing a precomputed mapping from trigger events to edge updates and transmitting the associated information to the decoding engine in a time-frame that coincides with the incoming syndrome data.

In the case of leakage-aware decoding, a trigger event involves measuring a leaked qubit (heralding). Each trigger is mapped to a set of edges that correspond to the error mechanisms that are made more likely by assuming the measured qubit was leaked since its last reset (Fig.~\ref{fig:wiggling}). Given a set of heralded leakage events, the adaptivity engine addresses its map to construct a total set of sensitive edges. Then, it sends each sensitive edge to the decoding engine to be pregrown.

\begin{figure*}[t!]
\centering
\includegraphics[scale=0.475]{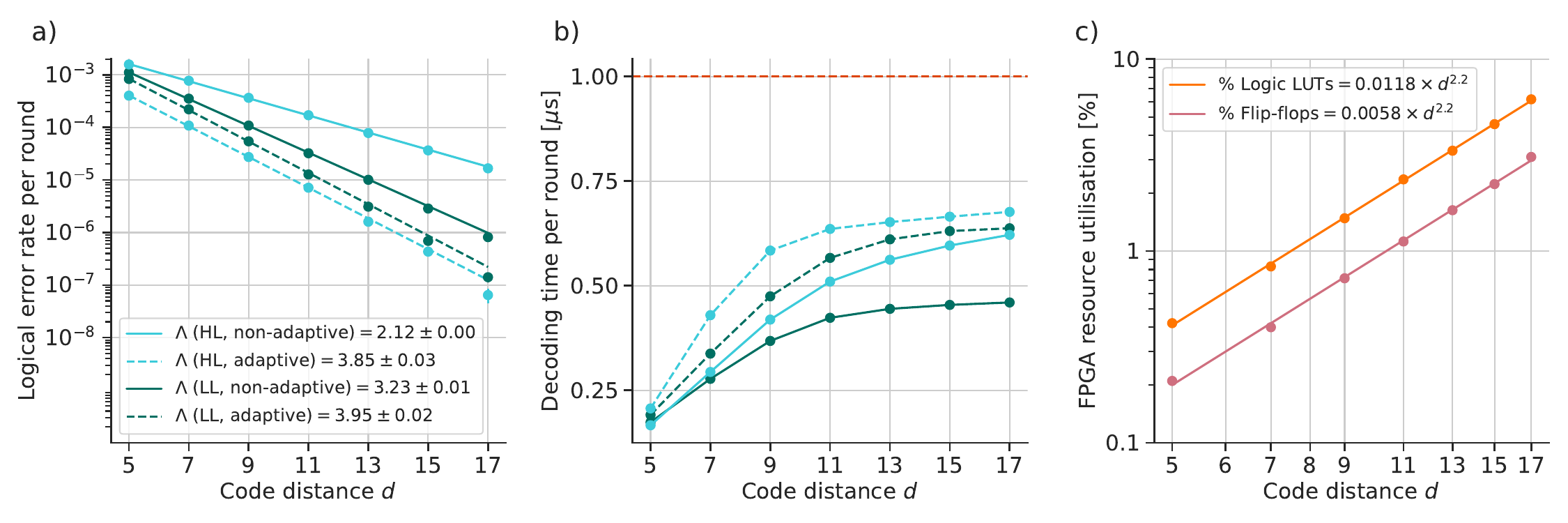}
\caption{
\textbf{Performance of LCD on a Xilinx Virtex Ultrascale+ VU19P FPGA \cite{vu19p}}. Each experiment uses a distance $d$ rotated planar surface code with patch wiggling and is sampled and decoded 10 million times. \textbf{(a)}
Logical error rate per round as a function of $d$. Each noise model: (LL) $p = 1\times10^{-3}$ and $p_l = 1\times10^{-4}$ (green); and (HL) $p = p_l = 5\times10^{-4}$ (blue), is decoded with non-adaptive (solid) and adaptive (dashed) decoding, resulting in different $\Lambda$. \textbf{(b)} Decoding time per round as function of $d$. The number of vertices per PE is $\lfloor {d/2} \rfloor$ and the operating frequency is 285MHz for all code distances, resulting in sublinear scaling with respect to $d$. We decode in under 1~$\mu$s per round (red line) on both noise models using non-adaptive or adaptive decoding up to at least distance $d=17$. \textbf{(c)} Log-log plot of the number of logic Look-Up-Tables (LUTs) and Flip-Flops (FFs) required by our decoder as a percentage of the total logic LUTs (4,085,760) and total FFs (8,171,520) available on the FPGA. At $d = 17$, we use around 6\% and 3\% of the available logic LUTs and FFs, respectively.}
\label{fig:fpga_performance_data}
\end{figure*}

For each possible heralded leakage event in the unit of logic that the decoder operates on, we precompute the corresponding adaptations by analysing the sensitivity of edges in the decoding graph to nearby leakage \cite{suchara_leakage_2014} (see also Methods). In practice, we do this by assuming the qubit could have leaked at any point since it was last reset, which adds high-probability depolarising channels on the qubits that interact with the leaked qubit.  From here we can compute which edge-like error mechanisms have become more likely given the leakage event under study by comparing the new circuit's error model to the baseline error model from the leakage-free circuit. We use Stim to generate error models for the circuits that drive our analysis \cite{gidney_stim_2021}.

\subsection{Performance}

Fig.~\ref{fig:fpga_performance_data} presents the performance of our decoder on a Xilinx Virtex Ultrascale+ VU19P FPGA.

We use a superconducting-inspired circuit-level noise model \cite{gidney_fault-tolerant_2021} and supplement it with stochastic leakage and relaxation channels. To simulate it, we use a modified version of Stim \cite{gidney_stim_2021} (see Methods). Our noise model has two variable parameters, $p$ and $p_l$, determining the strength of Pauli noise and leakage that follow a 2-qubit gate. We study two regimes:

\begin{itemize}
    \item Low leakage (LL): $p = 1\times10^{-3}$ and $p_l = 1\times10^{-4}$
    \item High leakage (HL): $p = p_l = 5\times10^{-4}$\,.
\end{itemize}

Complete details of our noise model can be found in Methods. These two noise regimes have a similar total amount of noise, but differ in the proportion of Pauli and leakage noise. These correspond to two possible futures for superconducting qubits: one where overall improvements in gate fidelities leave leakage as a dominant error (HL); and another in which leakage remains a substantially less common error than Pauli channels (LL).

Fig.~\ref{fig:fpga_performance_data}~(a) shows that, for the leakage-dominated noise model (HL, in light blue), adaptive decoding enables an improvement in $\Lambda$ from 2.12 to 3.85. To translate this accuracy gain into qubit savings, consider a target error probability of $10^{-6}$ for a $d\times d\times d$ window---a proxy of a fundamental unit of logic in lattice surgery \cite{horsman_surface_2012}. Non-adaptive decoding would require $d=33$, whereas adaptive decoding requires only $d=17$. Since in the surface code the number of physical qubits scales with $d^2$, adaptive decoding enables a near $75\%$ reduction in the number of physical qubits required compared to standard non-adaptive decoding. It has long been known that, without adaptivity, leakage events can effectively halve the distance of the code \cite{fowler_coping_2013}, and so our results show that adaptivity can reverse this phenomenon, making leakage errors no more pernicious than Pauli errors (see Supplementary Information for a more detailed discussion). For the noise model where leakage is less dominant (LL, in green), adaptivity still yields a 18\% reduction in the QEC footprint by improving $\Lambda$ from 3.23 to 3.95, requiring respectively $d=21$ and $d=19$ for $d\times d\times d$ at $10^{-6}$.

Fig.~\ref{fig:fpga_performance_data}~(b) shows timing results for our decoder.  We plot decoding time per round, obtained by dividing by $d$ the time required to decode a $d\times d\times d$ window.  In both noise regimes, we decode in less than 1~$\mu$s per round up to $d=17$ with and without adaptivity. The scaling of the decoding time per round is sublinear in $d$, since the number of vertices per PE is $\lfloor {d/2} \rfloor$ for all $d$. We do not measure the execution time of the adaptivity engine, since this can be highly optimised and pipelined alongside the decoder.

With coarse-grained parallelism, our decoder uses modest resources on a high-end FPGA, see Fig.~\ref{fig:fpga_performance_data}~(c) (for utilisation data on a mid-range FPGA, see Methods). The resource utilisation scales roughly as $d^2$, as expected given our vertex to PE mapping of the $d\times d \times d$ experiment.

A reduction in the maximum operating frequency of the FPGA implementations of LCD is observed as the code distance increases (see Methods). This is due to the increased routing complexity that results from the increase in resource utilisation. Bigger FPGAs can relax routing complexities and unlock higher operating frequencies. Alternatively, higher operating frequencies could be achieved via an ASIC implementation where logic placement can be determined more efficiently. An ASIC implementation of LCD is not explored at this point in time. However, LCD is well suited for ASIC given its parallel architecture.

\section{Discussion}

We have introduced an FPGA implementation of the Local Clustering Decoder, and have studied its performance under a circuit-level noise model with leakage.  We have presented results on the accuracy and speed of our decoder, as well as other metrics such as FPGA resource utilisation. Resource utilisation is a proxy for cost of and power consumed by the decoder, which is especially critical in cryogenic setups. Previous hardware decoders for circuit-level noise models have optimised for a variety of metrics. For example, Collision Clustering (CC) \cite{barber_real-time_2023} uses a serial architecture to prioritise resource utilisation over speed. Helios~\cite{liyanage_scalable_2023, liyanage_fpga-based_2024} can decode large surface codes in under 1~$\mu$s per QEC round due to its fine-grained parallel architecture, but its computational resource requirements scale more steeply in $d$ compared with CC: $O(d^3)$ vs $O(1)$. Our decoder uses a coarse-grained parallel architecture that interpolates between these extremes, resulting in computational resources scaling like $O(d^2)$ whilst maintaining high throughput.

The speed of a decoder impacts its throughput and latency. Throughput measures the rate at which syndromes can be consumed. To avoid the backlog problem \cite{terhal_quantum_2015}, throughput needs to be no smaller than the rate at which syndromes are generated---around once per $\mu$s in superconducting qubits.  Latency is a key component of the response time of the QEC system. The response time contributes directly to the logical clock rate of an error-corrected quantum computer. 

Low decoding latencies are necessary for fast error-corrected quantum computers needing conditional logic for magic-state teleportation. In the architecture of e.g.~\cite{litinski_game_2019}, teleportation is done with lattice-surgery measurements across several logical qubits, taking $O(d)$ rounds, followed by a reaction conditioned on the outcome of that decoded measurement \cite{bravyi_universal_2005, bravyi_magic-state_2012} (see however \cite{fowler_time-optimal_2013} for a time-optimal strategy for logical computation at a cost in the number of ``routing'' logical qubits, and \cite{litinski_game_2019, gidney_flexible_2019} for improvements to this strategy). Extracting $d$ rounds of syndromes takes around $d$~$\mu$s. This, together with the response time of the QEC system, determines the time that it takes to complete the teleportation of a magic state, setting the logical clock rate in such a lattice-surgery architecture. When pursuing fast logical clocks, a sensible target for the decoder latency is about $d$~$\mu$s for an $O(d)$ window, or 1~$\mu$s/round. If the decoder is faster, the logical clock rate is dominated by the duration of syndrome extraction, and that is what should be accelerated.

It has recently been understood that windowing and parallelism can deliver high throughput \cite{wu_fusion_2023, skoric_parallel_2023, tan_scalable_2023, bombin_modular_2023}---even slow decoders can avoid the backlog problem, given enough resources to parallelise over. Slow windowed decoders will, however, be slow at decoding their last window, incurring large latencies.  As we have seen in Fig.~\ref{fig:fpga_performance_data}~(b), LCD decodes up to distance $d=17$ in under 1~$\mu$s per round, meeting our latency targets. Further, meeting throughput requirements without parallelisation will simplify scaling into the large-scale fault-tolerant regime.

Recent work by Google \cite{acharya_quantum_2024} uses a specialized workstation to decode up to one million syndrome extraction rounds in real time with a parallelised algorithm similar to Fusion Blossom \cite{wu_fusion_2023}. The reported latency is approximately constant regardless of the number of rounds, at around 64~$\mu$s at $d=5$. Riverlane and Rigetti report a full QEC system response time of under 10~$\mu$s for a window of 9 rounds of syndrome extraction in a small $2\times 2$ quantum stability experiment \cite{caune_demonstrating_2024}. This was achieved by integrating the hardware CC decoder into Rigetti's superconducting device. We expect a similar latency in our decoder, and will measure the latency of a windowed version of LCD in a streaming scenario in future work.

We have seen how the adaptivity feature of LCD can greatly increase the decoder's accuracy under a noise model with leakage.  Accuracy can be further improved by adding weights to the decoding graph. 
This will require more FPGA resources compared to the unweighted implementation we have shown here, but given the modest resources that our current implementation utilises (around $6\%$ at $d=17$), we do not expect any significant obstacles in adding weights. When weights are available, we can consider weighted adaptivity maps, that further increase accuracy.  Examples of decoding strategies that will benefit from weighted adaptivity include two-pass correlated decoding \cite{fowler_optimal_2013}, soft information \cite{pattison_improved_2021, ali_reducing_2024}, and indeed weighted leakage-aware decoding \cite{suchara_leakage_2014}. While the present work is focussed on superconducting-inspired noise models, adaptive decoding  will also boost accuracy in AMO platforms, where leakage and atom loss are dominant noise channels \cite{evered_high-fidelity_2023}.

Looking beyond memory and towards computation, a scalable decoder must support fast reconfiguration of the decoding graph between windows. In certain scenarios, this can be achieved simply by switching to a sub graph of the compile-time graph where a subset of the vertices are turned off. Note that if a vertex is turned off, each of its incident edges are also turned off. This is a highly efficient procedure that can be run between adjacent windows of a long-running experiment without significantly impacting throughput and latency.

Where existing hardware decoders assume a regular structure to the decoding graph, a large class of graphs can be mapped onto LCD's PE array, simply by changing the connections between PEs. We have seen one example of this flexibility with the wiggling circuits which, despite having non-standard decoding graphs (with alternating orientation of hook edges), can be handled by LCD without any dedicated optimisations. In the past, accuracy and flexibility have been associated with software decoders. The present work demonstrates that hardware decoders can be fast, accurate, and flexible.

\section{Methods}

\subsection{Decoding algorithm}

LCD's decoding engine operates on a decoding graph as defined by its set of vertices $V$ and edges $E$, and takes two inputs: (1) a syndrome $S \subset V$; and (2) a set of pre-grown edges $P \subset E$. An edge is pre-grown if the probability of it being an error is much higher than other edges. The adaptivity engine is responsible for determining which edges to pre-grow based on trigger events, such as measuring a leaked qubit (heralding).

Clustering algorithms based on UF start by initialising a singleton cluster around each defect $u \in S$. A cluster is a set of vertices in the decoding graph whose parity is determined by the number of defects it contains. Then, each odd parity cluster that is not touching one of the open boundaries of the code grows by incrementing a support variable on each of its boundary edges, where a boundary edge is an edge with one endpoint in the cluster and the other endpoint outside the cluster. Subsequently, if two clusters share a new fully-grown edge, they merge into one bigger cluster. A fully-grown edge is an edge whose support is equal to its weight. These steps are repeated until all clusters have even parity or touch one of the open boundaries. The final set of clusters can be used to find a correction $C$ for the syndrome $S$ by peeling a spanning forest constructed from it, and the logical observable should be flipped post-measurement if $|C \cap L| \equiv 1 \pmod{2}$, where $L$ is the set of edges defining the logical observable.

We use a distributed version of this algorithm, akin to those described in \cite{liyanage_scalable_2023} and \cite{chan_actis_2023}. Here, each vertex is given a set of data fields that encode information about itself and its cluster. These data fields, which are modified by the PEs assigned to the vertex and its neighbours as the clusters grow and merge throughout the decoding graph, are as follows. For a given vertex $u$:

\begin{itemize}
\item \texttt{cindex} is an integer that identifies its cluster, and is equal to the lowest index of all vertices in the cluster. We say $u$ is a root vertex if $u.\texttt{cindex} = u.\texttt{index}$. Since all vertices start in singleton clusters, all vertices are initially root vertices.
\item \texttt{parent} is a pointer to the parent of $u$ in the tree of its cluster. Root vertices are their own parents.
\item \texttt{radius} is an integer that is equal to the radius of a circle centred at $u$. Initially, $u.\texttt{radius} = 0$.
\item \texttt{defect} is a bit that is equal to 1 if $u \in S$.
\item \texttt{parity} is a bit that is equal to the sum modulo 2 of the parities of $u$ and each of its children, or 0 if the vertex is connected by a fully-grown edge to a virtual boundary vertex. If $u$ is a root vertex, it is equal to the parity of its cluster. Initially, $u.\texttt{parity} = u.\texttt{defect}$.
\item \texttt{active} is a bit that is equal to 1 if $u$ is in an odd cluster that requires growth, i.e., active vertices grow. Initially, $u.\texttt{active} = u.\texttt{defect}$.
\item \texttt{busy} is a bit that is equal to 1 if any of the above data fields change and the current stage needs to be rerun for $u$'s neighbours to register that change. Initially, $u.\texttt{busy} = 0$.
\end{itemize}

Additionally, each vertex is given a set of pointers to its neighbours and a subset thereof that are accessible. We say a neighbour $v$ of a vertex $u$ is accessible if either: (1) the sum of their radii is greater than or equal to the weight of the edge between them; or (2) the edge between them is pregrown by the adaptivity engine. Each vertex is also given a read-only attribute $w_{\textrm{max}}$ that represents the maximum weight of all edges incident to it, which is used as a threshold for its radius. If the decoding graph is unweighted, $w_{\textrm{max}}$ must be hardcoded to 2 for all vertices, ensuring growth steps can proceed without expensive floating point calculations. Furthermore, all $w_{uv}$ where $(u, v) \in E$ must also be hardcoded to 2. The results in this paper are based on unweighted decoding graphs.  

\begin{figure}[t!]
\centering
\includegraphics[width=0.4\textwidth]{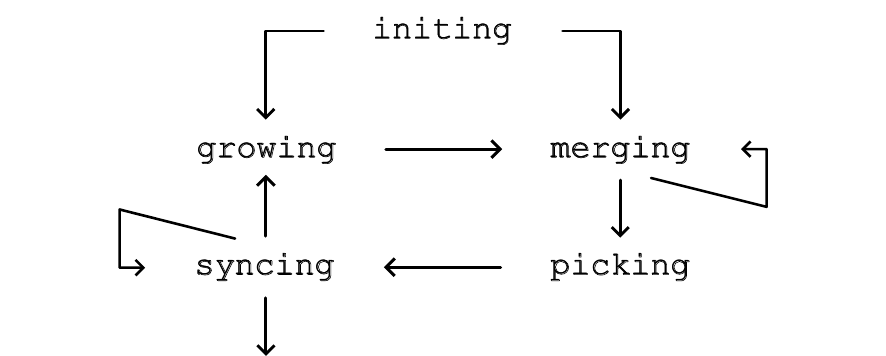}
\caption{\textbf{State machine}}
\label{fig:flowchart}
\end{figure}

\begin{algobox}[Controller FSM]
\begin{minipage}{\linewidth}
\begin{algorithmic}
  \Procedure{AdavanceController}{}
    \If{$(\text{not}~u.\texttt{busy})~\forall u \in V$}
      \If{$\texttt{stage} = \texttt{initing}$}
         \If{$|P| \neq 0$}
           \State $\texttt{stage} \gets \texttt{merging}$
         \Else
           \State $\texttt{stage} \gets \texttt{growing}$
         \EndIf
      \Else
        \If{$\texttt{stage} = \texttt{syncing}$}
          \If{$(\text{not}~u.\texttt{active})~\forall u \in V$}
            \State $\texttt{stage} \gets \texttt{exiting}$
          \Else
            \State $\texttt{stage} \gets \texttt{growing}$
          \EndIf
        \Else
            \State $\texttt{stage} \gets \text{next stage}$
        \EndIf
      \EndIf
    \EndIf
  \EndProcedure
\end{algorithmic}
\end{minipage} \\
\label{box:controller_fsm}
\end{algobox}

The state machine consists of an initialisation stage followed by four core stages (Fig.~\ref{fig:flowchart}). The core stages run cyclically in series until all clusters are neutralized, i.e., have even parity or touch one of the open boundaries. After the initialisation stage, a pre-clustering sequence must be run if there exists at least one pregrown edge, i.e., $|P| \neq 0$, which is achieved by starting the main loop in the merging instead of the growing stage. Note that the merging and syncing stages may rerun if any vertex exits in a busy state. The loop terminates after the syncing stage if there are no active vertices. This logic is captured by the \textsc{AdvanceController} procedure in Box~\ref{box:controller_fsm}.

\begin{algobox}[Stage kernels]
\begin{minipage}{\linewidth}
\begin{algorithmic}
  \Procedure{Growing}{$u$}
    \If{$(u.\texttt{active})~\text{and}~(u.\texttt{radius} \leq u.\texttt{w\textsubscript{max}})$}
      \State $u.\texttt{radius} \gets u.\texttt{radius} + 1$
    \EndIf
  \EndProcedure
  \State
  \Procedure{Merging}{$u$}
    \ForAll{$v \in u.\texttt{accessibles}$}
      \If{$(u.\texttt{cindex} > v.\texttt{cindex})$}
        \State $u.\texttt{cindex} \gets v.\texttt{cindex}$
        \State $u.\texttt{parent} \gets v$
        \State $u.\texttt{busy} \gets \texttt{true}$
      \EndIf
    \If{$(u.\texttt{parity})~\text{and}~(u.\texttt{parent} \neq u)$}
      \State $\text{Flip}~u.\texttt{parent}.\texttt{parity}$
      \State $u.\texttt{parity} \gets \texttt{false}$
      \State $u.\texttt{busy} \gets \texttt{true}$
    \EndIf
    \EndFor
  \EndProcedure
  \State
  \Procedure{Picking}{$u$}
    \State $u.\texttt{active} \gets u.\texttt{parity}$
  \EndProcedure
  \State
  \Procedure{Syncing}{$u$}
    \State $u.\texttt{busy} \gets \exists v \in u.\texttt{accessibles}: v.\texttt{active}$
    \State $u.\texttt{busy} \gets u.\texttt{busy}~\text{and}~(\text{not}~u.\texttt{active})$
    \State $u.\texttt{active} \gets u.\texttt{active}~\text{or}~u.\texttt{busy}$
  \EndProcedure
\end{algorithmic}
\end{minipage} \\
\label{box:stage_kernels}
\end{algobox}

In the merging and syncing stages, each part must execute its PEs in serial, since inter-PE communication is required. However, in the growing and picking stages, each part may execute its PEs in parallel, since no inter-PE communication is required. In either case, PEs apply a stage by executing the stage kernel on the subset of their vertices that are in a cluster, where a vertex is in a cluster if it is either: (1) a defect; or (2) incident to a fully-grown edge. This selective execution schedule leads to more efficient decoding, both in terms of power and performance, particularly at sub-threshold, where the number of vertices in a cluster is much less than the number of vertices in the decoding graph \cite{griffiths_union-find_2024}.

\begin{figure*}[t!]
\centering
\includegraphics[scale=0.30]{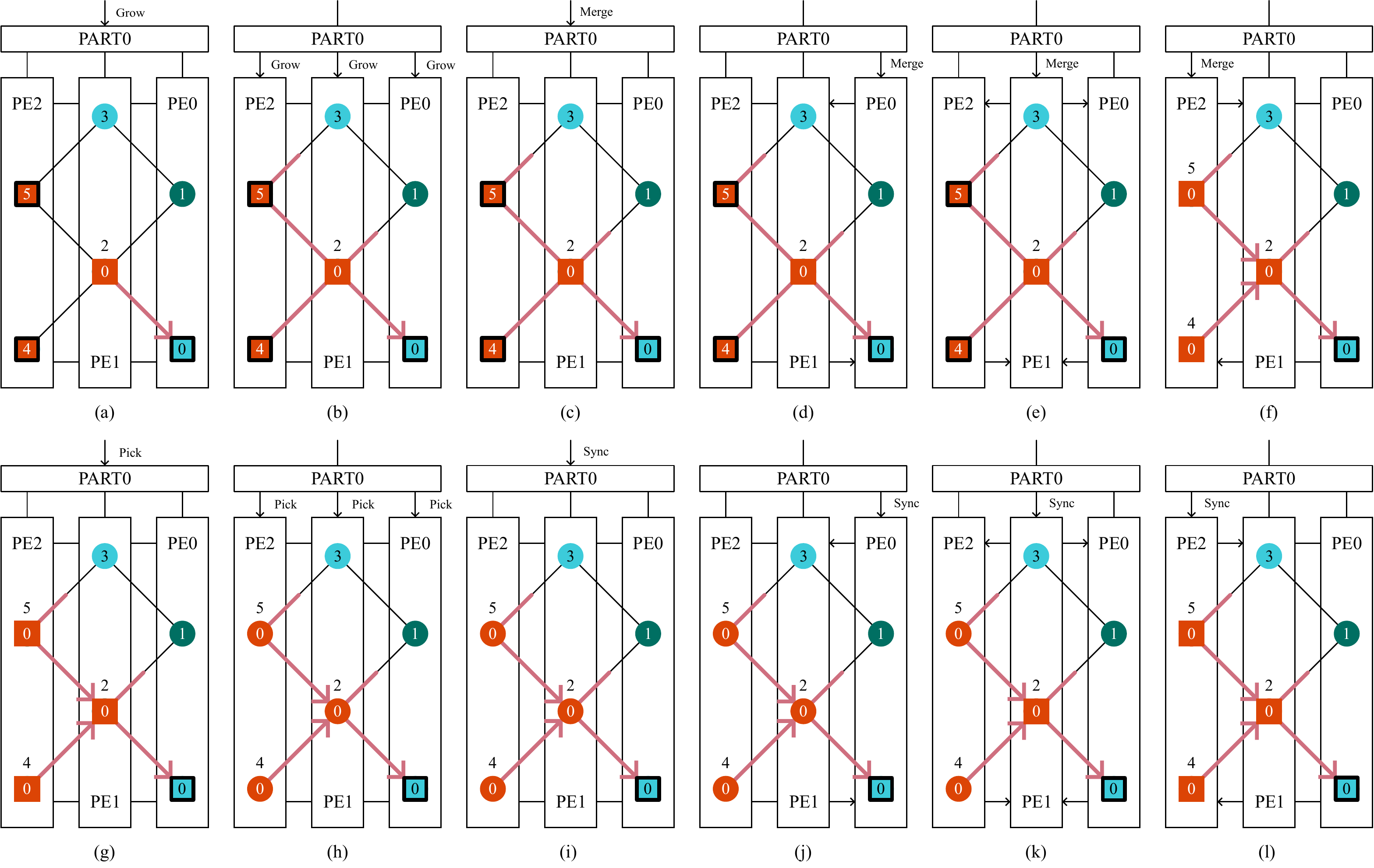}
\caption{\textbf{Pass through the finite state machine (FSM)}. Bulk vertices are green; boundary-adjacent blue. Red vertices are defects. Square vertices are active; circular inactive. Black-outlined vertices have odd parity. Labels inside vertices are cluster indices; labels above, vertex indices. The decoding graph is unweighted. If the radii of the vertices on the endpoints of an edge sum to 2 or the edge is a pre-grown edge, we say it is fully-grown and colour it pink. Parenthood relationships are represented with arrows. There are three odd clusters, two singleton clusters ($\{4\}$ and $\{5\}$) and one pre-cluster ($\{0, 2\}$) \textbf{(a)} Part 0 enters the \texttt{growing} stage. \textbf{(b)} Active vertices increase their radii by one. Since vertices 0, 2, 4 and 5 are active, this results in fully-grown edges $\{(2, 4), (2, 5)\}$. \textbf{(c)} Part 0 enters the \texttt{merging} stage. \textbf{(d - f)} Each vertex identifies the neighbour connected to it by a fully-grown edge with the lowest cluster index. Then, if the cluster index of the neighbour is less than that of the vertex, the vertex adopts it as its own and begins pointing at the neighbour. Lastly, child vertices with odd parity, flip the parities of their parents, and make their own even. In part \textbf{(f)}, vertices 4 and 5 adopt cluster index 0 from vertex 2 and begin pointing at it. As such, the parities of vertices 2, 4 and 5 flip twice, once and once, respectively, making them all even. At the end of a merge sequence, all vertices, excluding root vertices, must have even parity. \textbf{(g)} Part 0 enters the \texttt{picking} stage. \textbf{(h)} Root vertices with odd parity become active. Since vertex 0 is the root of the odd cluster $\{0, 2, 4, 5\}$, it becomes active. The other vertices in the cluster, i.e., 2, 4 and 5, become inactive. \textbf{(i)} Part 0 enters the \texttt{syncing} stage. \textbf{(j - l)} Each vertex connected by a fully-grown edge to a neighbour that is active, becomes active. In part \textbf{(k)}, vertex 2 becomes active through vertex 0. In part \textbf{(l)}, vertices 4 and 5 become active through vertex 2.}
\label{fig:fsm}
\end{figure*}

The stage kernels are described with pseudo code in Box~\ref{box:stage_kernels} and by example in Fig.~\ref{fig:fsm}. Note that we use a radius on each vertex instead of a support on each edge, with the support of an edge being computed at runtime by summing the radii of the vertices on its endpoints. This is more space efficient, since the number of edges in a decoding graph is typically an order of magnitude greater than the number of vertices. For example, the maximum degree of a circuit-level decoding graph for the surface code is 12, since a vertex can have up to 4 neighbours in the same round and 8 in adjacent rounds. 

\subsection{Runtime}

As in previous implementations, the runtime of our decoder is dominated by the merging and syncing stages; which require $O(w)$ time steps for a cluster of diameter $w$, since the time to send information up or down the tree of a cluster depends on its diameter. In the worst case, we have $w = d^3$---the cluster forms a winding path of length $O(d^3)$ \cite{chan_actis_2023}. However, below threshold the mean cluster diameter is much smaller than $d^3$ and depends only on the physical error rate $p$, i.e., it is invariant with $d$ \cite{griffiths_union-find_2024}. In practice, this leads to a decoding time that scales sub linearly with $d$ (Fig.~\ref{fig:fpga_performance_data}~(b)).

\subsection{Hardware implementation details}

LCD was implemented using the SystemVerilog hardware description language (HDL). The PE, part and controller modules in LCD were manually written, and the inter-PE NoC module was auto-generated using an in-house utility. The source code was kept implementation-agnostic, so it can be used for both FPGA and ASIC implementations. For generating the performance data, LCD was implemented on a high-end Xilinx Virtex Ultrascale+ VU19P FPGA~\cite{vu19p} and a mid-range Xilinx Zynq UltraScale+ ZCU111 FPGA~\cite{zcu111}.

The hardware design was optimized for high performance and modest resource utilization. Table-\ref{tab:vu19p-utilization} \& Table-\ref{tab:zcu111-utilization} show the maximum operating frequency and resource utilisation of the FPGA implementations of LCD across a range of surface code distances. The main FPGA resources~\cite{ug901} considered are Lookup Tables (LUTs), used for logic, and Flip-Flops (FFs), used for storage.

The FPGA implementations of LCD aim for the highest possible operating frequency. However, as resource utilisation increases, routing delays introduced by the Place and Route (P$\&$R) tool become more pronounced as it tries to balance resources available in a given area with the provided timing constraints. Since FPGAs have limited resources, this imposes an upper bound on the maximum code distance that a given FPGA can support whilst maintaining MHz decoding rates. This would not be an issue for ASIC implementations since the design can be laid out with finer control. Table-\ref{tab:hw-decoding-time} lists the decoding time per round for each configuration as a function of the maximum operating frequency.

\begin{table*}[t!]
\setlength{\tabcolsep}{6pt}
\centering
\begin{tabular}{clccccccc}
\toprule
\multicolumn{2}{c}{\multirow{2}{*}{\textbf{XCVU19P}}} & \multicolumn{7}{c}{\textbf{Code Distance}} \\
\multicolumn{1}{c}{} & & \textbf{5} & \textbf{7} & \textbf{9} & \textbf{11} & \textbf{13} & \textbf{15} & \textbf{17}\\
\toprule
\multicolumn{2}{c}{\textbf{Maximum Frequency}} & \multirow{2}{*}{400} & \multirow{2}{*}{384} & \multirow{2}{*}{370} & \multirow{2}{*}{333} & \multirow{2}{*}{312} & \multirow{2}{*}{303} & \multirow{2}{*}{285} \\
\multicolumn{2}{c}{\textbf{(MHz)}} & & & & & & & \\
\toprule
\multirow{11}{*}{\textbf{Site Type}} & \multirow{2}{*}{Logic LUTs} & 17003 & 34060 & 60640 & 96438 & 135942 & 186996 & 251963 \\
\cmidrule[\heavyrulewidth]{3-9}
\multicolumn{2}{c}{} & 0.42\% & 0.83\% & 1.48\% & 2.36\% & 3.33\% & 4.58\% & 6.17\% \\
\cmidrule[\heavyrulewidth]{2-9}
& \multirow{2}{*}{LUTRAM} & 520 & 1024 & 2136 & 3000 & 4520 & 5808 & 7872\\
\cmidrule[\heavyrulewidth]{3-9}
\multicolumn{2}{c}{} & 0.05\% & 0.11\% & 0.22\% & 0.31\% & 0.47\% & 0.61\% & 0.82\%\\
\cmidrule[\heavyrulewidth]{2-9}
& \multirow{2}{*}{SRLs} & 1240 & 2220 & 4280 & 6186 & 8430 & 11037 & 16554\\
\cmidrule[\heavyrulewidth]{3-9}
\multicolumn{2}{c}{} & 0.13\% & 0.23\% & 0.45\% & 0.65\% & 0.88\% & 1.15\% & 1.73\%\\
\cmidrule[\heavyrulewidth]{2-9}
& \multirow{2}{*}{FFs} & 16817 & 32983 & 58905 & 91156 & 133532 & 181949 & 252736 \\
\cmidrule[\heavyrulewidth]{3-9}
\multicolumn{2}{c}{} & 0.21\% & 0.40\% & 0.72\% & 1.12\% & 1.63\% & 2.23\% & 3.09\%\\
\bottomrule

\end{tabular}
\caption{\label{tab:vu19p-utilization}LCD hardware implementation statistics on Xilinx XCVU19P FPGA.}
\end{table*}

\begin{table*}[t!]
\setlength{\tabcolsep}{6pt}
\centering
\begin{tabular}{clccccccc}
\toprule
\multicolumn{2}{c}{\multirow{2}{*}{\textbf{ZCU111}}} & \multicolumn{7}{c}{\textbf{Code Distance}} \\
\multicolumn{1}{c}{} & & \textbf{5} & \textbf{7} & \textbf{9} & \textbf{11} & \textbf{13} & \textbf{15} & \textbf{17}\\
\toprule
\multicolumn{2}{c}{\textbf{Maximum Frequency}} & \multirow{2}{*}{400} & \multirow{2}{*}{370} & \multirow{2}{*}{370} & \multirow{2}{*}{303} & \multirow{2}{*}{270} & \multirow{2}{*}{263} & \multirow{2}{*}{235} \\
\multicolumn{2}{c}{\textbf{(MHz)}} & & & & & & & \\
\toprule
\multirow{11}{*}{\textbf{Site Type}} & \multirow{2}{*}{Logic LUTs} & 17319 & 33240 & 60461 & 94657 & 134994 & 184182 & 245217 \\
\cmidrule[\heavyrulewidth]{3-9}
\multicolumn{2}{c}{} & 4.07\% & 7.82\% & 14.22\% & 22.26\% & 31.74\% & 43.31\% & 57.66\%\\
\cmidrule[\heavyrulewidth]{2-9}
& \multirow{2}{*}{LUTRAM} & 520 & 1024 & 2136 & 3000 & 4520 & 5808 & 7872 \\
\cmidrule[\heavyrulewidth]{3-9}
\multicolumn{2}{c}{} & 0.24\% & 0.48\% & 1.00\% & 1.40\% & 2.12\% & 2.72\% & 3.69\%\\
\cmidrule[\heavyrulewidth]{2-9}
& \multirow{2}{*}{SRLs} & 1240 & 2221 & 4280 & 6186 & 8430 & 11036 & 16554\\
\cmidrule[\heavyrulewidth]{3-9}
\multicolumn{2}{c}{} & 0.58\% & 1.04\% & 2.00\% & 2.90\% & 3.95\% & 5.17\% & 7.75\%\\
\cmidrule[\heavyrulewidth]{2-9}
& \multirow{2}{*}{FFs} & 16775 & 32905 & 59177 & 91017 & 133521 & 182146 & 253006 \\
\cmidrule[\heavyrulewidth]{3-9}
\multicolumn{2}{c}{} & 1.97\% & 3.87\% & 6.96\% & 10.70\% & 15.70\% & 21.41\% & 29.75\%\\
\bottomrule

\end{tabular}
\caption{\label{tab:zcu111-utilization}LCD hardware implementation statistics on Xilinx ZCU111 FPGA.}
\end{table*}

\begin{table*}[t!]
\setlength{\tabcolsep}{6pt}
\centering
\begin{tabular}{cccccccccc}
\toprule
\multicolumn{10}{c}{\textbf{Decoding Time Per Round ($\mu$s)}} \\
\toprule
\multirow{8}{*}{\textbf{XCVU19P}} & \multirow{3}{*}{\textbf{Decoding Type}} & \multirow{2}{*}{\textbf{Noise}} & \multicolumn{7}{c}{\textbf{Code Distance (Maximum Frequency, MHz)}} \\
\cmidrule[\heavyrulewidth]{4-10}
& & \multirow{2}{*}{\textbf{Regime}} & \textbf{5} & \textbf{7} & \textbf{9} & \textbf{11} & \textbf{13} & \textbf{15} & \textbf{17}\\
& & & 400MHz & 384MHz & 370MHz & 333MHz & 312MHz & 303MHz & 285MHz\\
\cmidrule[\heavyrulewidth]{2-10}
& Non-adaptive & LL & 0.124	& 0.206	& 0.283	& 0.362	& 0.406 & 0.427 & 0.460 \\
& Non-adaptive & HL & 0.118 & 0.218 &	0.323 &	0.436 & 0.513 &	0.561 &	0.622 \\
& Adaptive & LL & 0.136 &	0.251 &	0.366 &	0.485 &	0.558 & 0.593 & 0.637 \\
& Adaptive & HL & 0.147 & 0.319 &	0.450 &	0.544 & 0.596 &	0.626 &	0.676 \\
\toprule
\multirow{8}{*}{\textbf{ZCU111}}& \multirow{3}{*}{\textbf{Decoding Type}} & \multirow{2}{*}{\textbf{Noise}} & \multicolumn{7}{c}{\textbf{Code Distance (Maximum Frequency, MHz)}} \\
\cmidrule[\heavyrulewidth]{4-10}
& &  \multirow{2}{*}{\textbf{Regime}} & \textbf{5} & \textbf{7} & \textbf{9} & \textbf{11} & \textbf{13} & \textbf{15} & \textbf{17}\\
& & & 400MHz & 370MHz & 370MHz & 303MHz & 270MHz & 263MHz & 235MHz\\
\cmidrule[\heavyrulewidth]{2-10}
& Non-adaptive & LL & 0.124 & 0.214 &	0.283 &	0.398 & 0.469 & 0.492 & 0.558 \\
& Non-adaptive & HL & 0.118 &	0.226 &	0.323 &	0.480 &	0.593 &	0.646 &	0.754 \\
& Adaptive & LL & 0.136 &	0.260 &	0.366 &	0.533 &	0.645 &	0.683 &	0.773 \\
& Adaptive & HL & 0.147 &	0.331 &	0.450 &	0.598 & 0.688 &	0.721 & 0.820 \\
\bottomrule
\end{tabular}
\caption{\label{tab:hw-decoding-time} LCD hardware decoding time per round statistics.}
\end{table*}

\subsection{Accuracy comparison with PyMatching}

\begin{figure*}[t!]
\centering
\includegraphics[scale=0.85]{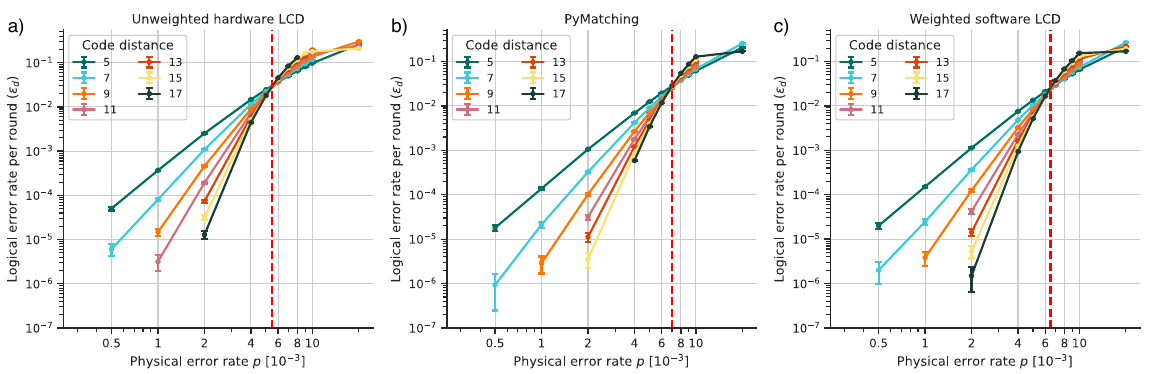}
\caption{
\textbf{Accuracy comparison} between \textbf{(a)} unweighted hardware LCD, \textbf{(b)} PyMatching and \textbf{(c)} weighted software LCD under a uniform circuit-level depolarising noise model without leakage and a standard syndrome extraction schedule.
}
\label{fig:threshold_plot}
\end{figure*}

In Fig.~\ref{fig:threshold_plot}, we provide an accuracy comparison between the unweighted hardware implementation of LCD and PyMatching \cite{higgott_pymatching_2022}, an open-source implementation of weighted MWPM. For reference, we also include results for a weighted software implementation of LCD. For all decoders, we plot the logical error rate per round on X basis quantum memory experiments using a variety of physical error rates and code distances. We use a uniform circuit-level depolarising noise model without leakage (SD6 in \cite{gidney_fault-tolerant_2021}) and a standard syndrome extraction schedule (in particular, without wiggling). We compute the logical error rate per round in accordance to Section VI.B of \cite{acharya_quantum_2024}:
\begin{equation}
\label{eq:lambda}
\epsilon_{d} = \frac{1}{2}\left(1 - \left(1 - 2P_{L}\right)^{1/t}\right),
\end{equation}
where $P_{L}$ is the logical error rate over $t$ rounds of error correction. In our experiments, $t = d$.

The threshold can be estimated by finding the point where the curves for each distance cross. This occurs at $p = 0.55\%$ for the unweighted hardware LCD and $p = 0.7\%$ for PyMatching. We note that this gap is bridged by using a weighted version of LCD implemented in software, with a threshold of around $p = 0.65\%$ and very similar accuracy to PyMatching.

\subsection{Modelling leakage with Pauli Frame tracking}

There are several existing approaches to modelling leakage in the context of QEC experiments, all of which are based on a trade-off between simulation accuracy and efficiency. To model leakage fully, density matrix simulations are required \cite{varbanov_leakage_2020, camps_leakage_2024}. While these are accurate, resources scale exponentially with the size of the system being simulated and therefore only experiments containing a handful of qubits can be modelled. At the other end of the scale, Pauli frame simulations such as those in \cite{fowler_coping_2013, suchara_leakage_2014, brown_critical_2020, acharya_suppressing_2023, vittal_eraser_2023, gu2024optimizingquantumerrorcorrection, gu2025} treat leakage classically and approximate two-qubit gate errors due to leakage stochastically. By making these approximations, large-scale simulations may be performed while accepting a loss of accuracy in modelling the system. Other approaches, such as tensor network simulations \cite{manabe_efficient_2023}, aim to strike a balanced trade-off.

We extend the Clifford circuit simulator Stim \cite{gidney_stim_2021} with a capability for stochastic simulation of leakage. Each qubit in the simulation is associated with a bit in an array that tracks the qubit moving in and out of the leaked state as the simulation progresses through time. Whether this bit is set to 1 (leaked) or 0 (not leaked, or ``sealed'') enables stochastic simulation of the damaging correlated errors that are characteristic of leakage within an efficient, highly parallelisable Pauli frame simulation architecture. With this, we implement the following functionality to enable our simulations:

\begin{table*}[th!]
\setlength{\tabcolsep}{18pt}
\begin{tabular}{@{}lll}

\toprule
\textbf{Operation} & \textbf{SI1000 with leakage} & \textbf{Strength}
\\
 \cmidrule{1-3}
CZ & Two qubit depolarising channel & $p$ \\
 & Leakage channel & $p_l$ \\
 & Relaxation channel & $p_l$ \\
One qubit Clifford gate & One qubit depolarising channel & $p/10$ \\
 & Relaxation channel & $p/5$ \\
Reset & One qubit depolarising channel & $2p$ \\
 & Leakage channel & $p_l$ \\
Measurement & Measurement error & $5p$ \\
 & Leakage heralding error & $5p$ \\
Idle & One qubit depolarising error & $p/10$ \\
 & Relaxation channel & $p/5$ \\
Resonator Idle & One qubit depolarising channel & $2p$ \\
 & Relaxation channel & $4p$ \\
\cmidrule{1-3}
\end{tabular}
\caption{\label{tab:noise_model} \textbf{Noise model simulated in this paper}. The core of this model replicates the SI1000, superconducting inspired, noise model \cite{gidney_fault-tolerant_2021}. However, we have extended the SI1000 specification with leakage and relaxation error channels.}

\end{table*}

\begin{enumerate}
    \item A one-qubit leakage channel, with a parameter determining the probability that the qubit will exit the computational basis states. This channel does not do anything to leaked qubits. When a sealed qubit leaks, the leakage register associated with the qubit is updated to 1. Furthermore, the qubit register gets fully depolarised by randomly applying $I$, $X$, $Y$ or $Z$ with equal probabilities. This is inconsequential at this point (as the qubit is leaked) but it ensures that, if the qubit relaxes or gets measured at a later point, it does so in a maximally mixed state.
    \item A one-qubit relaxation channel, with a parameter determining the probability that a leaked target qubit will return to the computational basis at a given point in time. This channel does not do anything to sealed qubits. When a leaked qubit returns to the computational subspace it is modelled as a maximally mixed state due to the prior depolarisation when it originally leaked. The leakage register associated with the qubit is updated to 0 to ensure no further Pauli errors are introduced when the newly sealed qubit undergoes further two qubit gates.
    \item Two-qubit gates involving leaked qubits introduce additional Pauli errors. That is, when a CZ gate is applied the qubit leakage registers are checked. If one of the qubits is in the leaked state, then the other qubit gets fully depolarised. This approximates the impact of leakage on a two-qubit gate. We note that fully depolarising the sealed qubit overwrites any propagation of Pauli errors introduced in the qubit register of the leaked qubit when that qubit was originally leaked (point 1.). 
    \item A reset gate unconditionally returns a qubit to the $\ket{0}$ state and the leakage register gets set to 0. This reflects the way unconditional resets extract leakage by returning leaked qubits to a sealed state.
    \item Measurements can herald leakage, reflecting how leaked states can be distinguished from sealed states through measurement \cite{miao_overcoming_2023, ali_reducing_2024}. Notably, heralding of leakage, as with all measurements, is imperfect. Heralding errors are modelled asymmetrically to reflect prevalent device physics which sees leaked states misclassified as sealed states (``false negatives'') dominating much rarer ``false positives'', where a sealed state is misclassified as leaked \cite{miao_overcoming_2023, varbanov_leakage_2020}. When heralding leakage, we parameterise heralding noise with a single probability determining the chance of a false negative heralding flip.
\end{enumerate}

\subsection{Leakage-aware decoding}

Prior work on leakage-aware decoding has focused on rebuilding the decoding graph in response to a set of herald events. This has been performed through a combination of reweighting existing edges and adding fundamentally new edges into the error model prior \cite{suchara_leakage_2014}. Our work differs in two notable ways.

Firstly, we show that leakage-aware decoding can be performed effectively in an unweighted decoding context. We handle leakage by pregrowing regions of LCD's decoding graph. This has the same effect as reweighting all edges affected by leakage to weight 0, assuming the decoder proceeds by merging all vertices along pregrown regions before beginning the grow-merge cycle that characterises standard UF. Pregrowing and preclustering minimal subsets of existing edges are effective approximations to the more granular reweighting approach of \cite{suchara_leakage_2014}. This is because once the decoder has certainty about leakage in a given location any updated edge weights will be highly diverged from the weights of edges that are unaffected by leakage.

Secondly, we do not need to add new edges to the graph when performing a graph transformation in response to a heralded leakage event. When we detect a herald, we know that the qubit must have leaked at some point in its history. Physically, this means two things: (A) errors in the system propagate differently (as the two-qubit gate no longer behaves as expected), and (B) new noise is introduced into the system. The modified error propagation (A) will lead to new hyperedges in the decoding hypergraph. However, the fully depolarising noise on the partner qubit (B) effectively overwrites this modified propagation, and we can therefore simply reweight existing edges. This observation allows us to streamline the process of performing real-time decoding and minimises the hardware complexity required to realise adaptivity on an FPGA.

\subsection{Noise model and native gate set}

Our noise model is an extension of the SI1000 superconducting-inspired noise model \cite{gidney_fault-tolerant_2021}. It is designed to capture key characteristics of superconducting qubits where measurements and resets make a dominant contribution to the time and error budget during a QEC experiment \cite{chen_exponential_2021}. For complete information on the noise model see Table-\ref{tab:noise_model}.

We study two different parametrisations of our noise model, where leakage is both a dominant and subdominant error mechanism by setting $p$ and $p_l$ as follows:

\begin{enumerate}
    \item \textbf{Low Leakage}:  $p = 10^{-3}$ and $ p_l = 10^{-4}$. Here we assume leakage is subdominant. That is, that the current contribution of leakage to overall error budgets in state of the art superconducting experiments \cite{acharya_suppressing_2023} can be sustained as overall qubit fidelities are improved.
    \item \textbf{High Leakage}: $p = p_l = 5\times10^{-4}$. A model in which improvements in qubit fidelities leave leakage as the dominant error.
\end{enumerate}

In likeness of SI1000, our relaxation parameters are informed by the dominant amount of time spent performing measurement gates during a QEC round implemented with superconducting qubits. For example, relaxation during idling is controlled by strength $p/5$ since the $T_1$ time that controls 
$|2\rangle\rightarrow |1\rangle$ decay for a harmonic oscillator is half that of the $T_1$ time that controls the $|1\rangle\rightarrow |0\rangle$ transition. Idling noise is thereby twice as strong when a qubit is leaked when compared against the $p/10$ channel that affects sealed idle qubits. The same argument motivates the $4p$ relaxation strength during resonator idle. Finally, the strength of a leakage heralding error matches that of a sealed measurement error. A heralding error is modeled asymmetrically as the probability of incorrectly classifying a leaked qubit as being in the sealed state.

The circuits simulated in this paper use the following native set of operations: H, CZ, single-qubit computational basis measurements (MZ) and single-qubit computational basis reset (RZ).

\section*{Data availability}

The data used in this study are available in the Zenodo database under accession code \href{https://doi.org/10.5281/zenodo.16982690}{10.5281/zenodo.16982690}.

\begin{acknowledgments}
We thank Earl Campbell and Neil Gillespie for input at various stages of this work.  We also thank them, together with Dan Browne and Maria Maragkou, for feedback on a draft of this paper. We are grateful to Ophelia Crawford for assistance with the wiggling circuits. We thank Steve Brierley and our colleagues at Riverlane for creating a stimulating environment for research.
\end{acknowledgments}

\section*{Author Contributions Statement}
A.B.Z.\ designed the decoding algorithm and high-level decoding architecture, and implemented the functional and performance models. M.L.T.\ designed and implemented the adaptivity features and the leakage-aware decoding strategy. The hardware micro-architecture was designed by A.Z.\ with input from A.B.Z., and the Register Transfer Level (RTL) description was written and implemented by A.Z.\ on FPGAs. C.T.\ contributed in making the hardware design feature rich and performant. J.C., G.P.G., M.P.S.\ and M.L.T.\ developed early concepts used by the leakage-aware decoder. M.L.T.\ wrote the leakage simulator with M.P.S., J.C., and G.P.G.\ giving input. A.B.Z.\ and C.T.\ designed and implemented the embedded software to drive the hardware decoder. The data in the paper was collected by A.B.Z.\ and A.Z. A.B.Z.\ and J.C.\ coordinated the project and wrote the plotting scripts. All authors contributed to the writing.

\section*{Competing interests Statement}
The authors declare no competing interests.

\end{document}